\documentclass[a4paper]{jpconf}
\usepackage{graphicx}
\begin{document}
\title{Driving and damping mechanisms in hybrid pressure-gravity modes pulsators}

\author{M A Dupret$^1$, A Miglio$^2$, J Montalban$^2$, A Grigahc\`ene$^3$ and A Noels$^2$}

\address{$^1$ Observatoire de Paris, LESIA, CNRS UMR 8109, 5 place J. Janssen, 92195 Meudon, France}
\address{$^2$ Institut d'Astrophysique et G\'eophysique, universit\'e de Li\`ege, Belgium}
\address{$^3$ CRAAG - Algiers Observatory BP 63 Bouzareah 16340, Algiers, Algeria}

\ead{MA.dupret@obspm.fr}

\begin{abstract}
We study the energetic aspects of hybrid pressure-gravity modes pulsations. The case of hybrid $\beta$~Cephei-SPB pulsators is considered with special attention. In addition to the already known sensitivity of the driving mechanism to the heavy elements mixture (mainly the iron abundance), we show that the characteristics of the propagation and evanescent regions play also a major role, determining the extension of the stable gap in the frequency domain between the unstable low order pressure and high order gravity  modes. Finally, we consider the case of hybrid $\delta$~Sct-$\gamma$~Dor pulsators.
\end{abstract}

\section{Introduction}

In the HR diagram, many families of pulsators appear as couple. In the hot part of the main sequence, we encounter the $\beta$~Cep stars with low radial order modes ($\nu\approx 3-8$ d$^{-1}$) having a mixed p-g mode nature (p-mode type in most of the envelope and g-mode type in the region just above the convective core). And just at their cool side the Slowly Pulsating B (SPB) stars with high radial order gravity modes ($\nu\approx 0.2-1$ d$^{-1}$, $n\approx 20-50$). The same occurs at lower temperature: we find the $\delta$~Sct stars with low radial order p-modes and just at their right the $\gamma$~Dor stars with high radial order g-modes; and at later evolution stages also we find the couple of sdB and Betsy stars. We concentrate here on the case of $\beta$~Cep and SPBs. Their modes are driven by a $\kappa$-mechansism operating in the Iron Opacity Bump (IOB) at $\log T\approx 5.3$. At first sight, the different pulsation periods and locations in the HR diagram  of these two families seem to take their origin in the location of this IOB. As $T_{\rm eff}$ decreases, it goes deeper into the star, entering progressively the domain of g-modes with larger periods. But recent observations complicate this simple scenario: hybrid stars with both types of oscillations at the same time are observed. The goal of this paper is to explain their existence and show some aspects of their high potential for asteroseismology. 

\section{Hybrid $\beta$~Cep - SPB stars}

\begin{figure}[h]
\begin{minipage}{18.5pc}
\includegraphics[width=19pc]{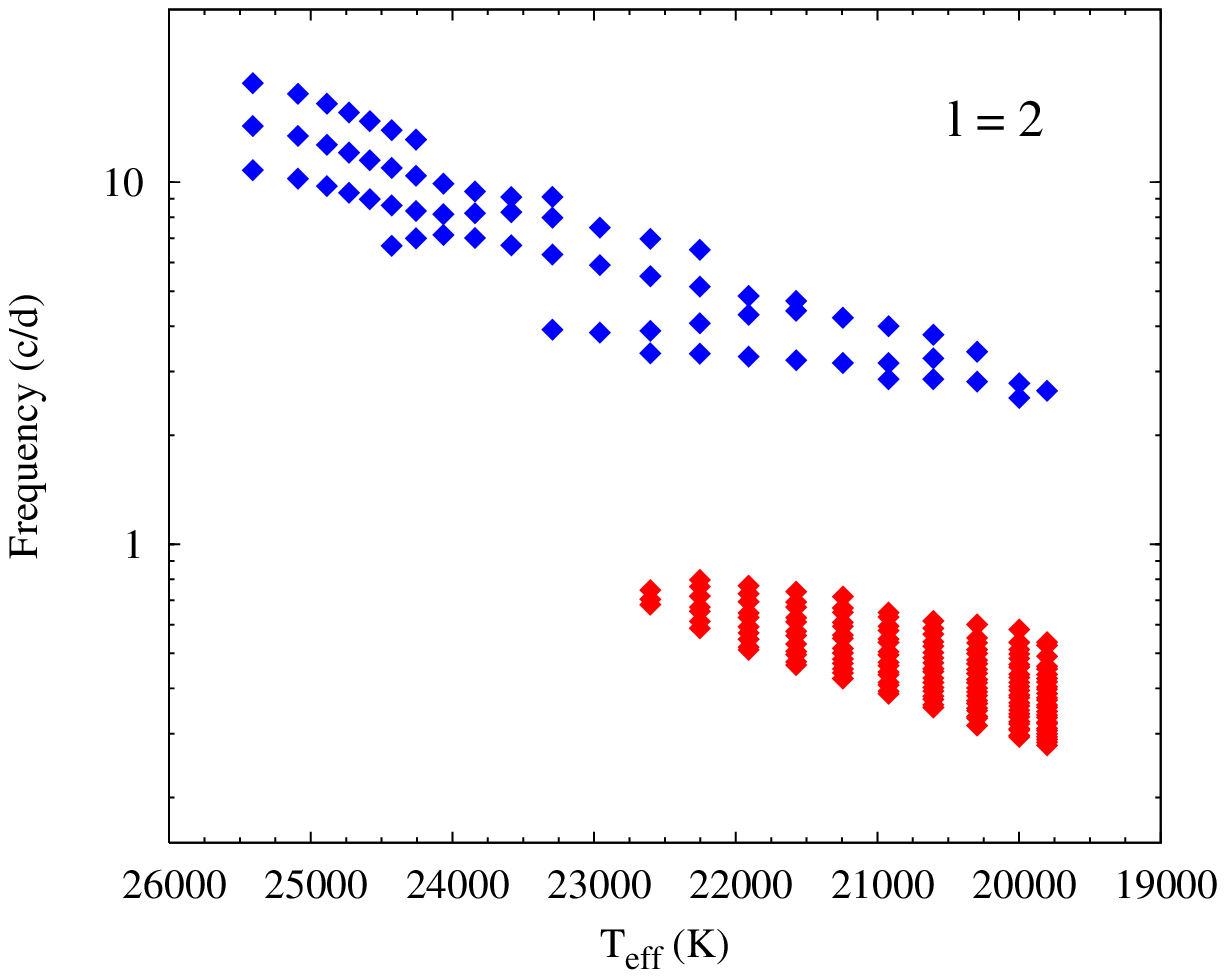}
\end{minipage}\hspace{0pc}%
\begin{minipage}{18.5pc}
\includegraphics[width=19pc]{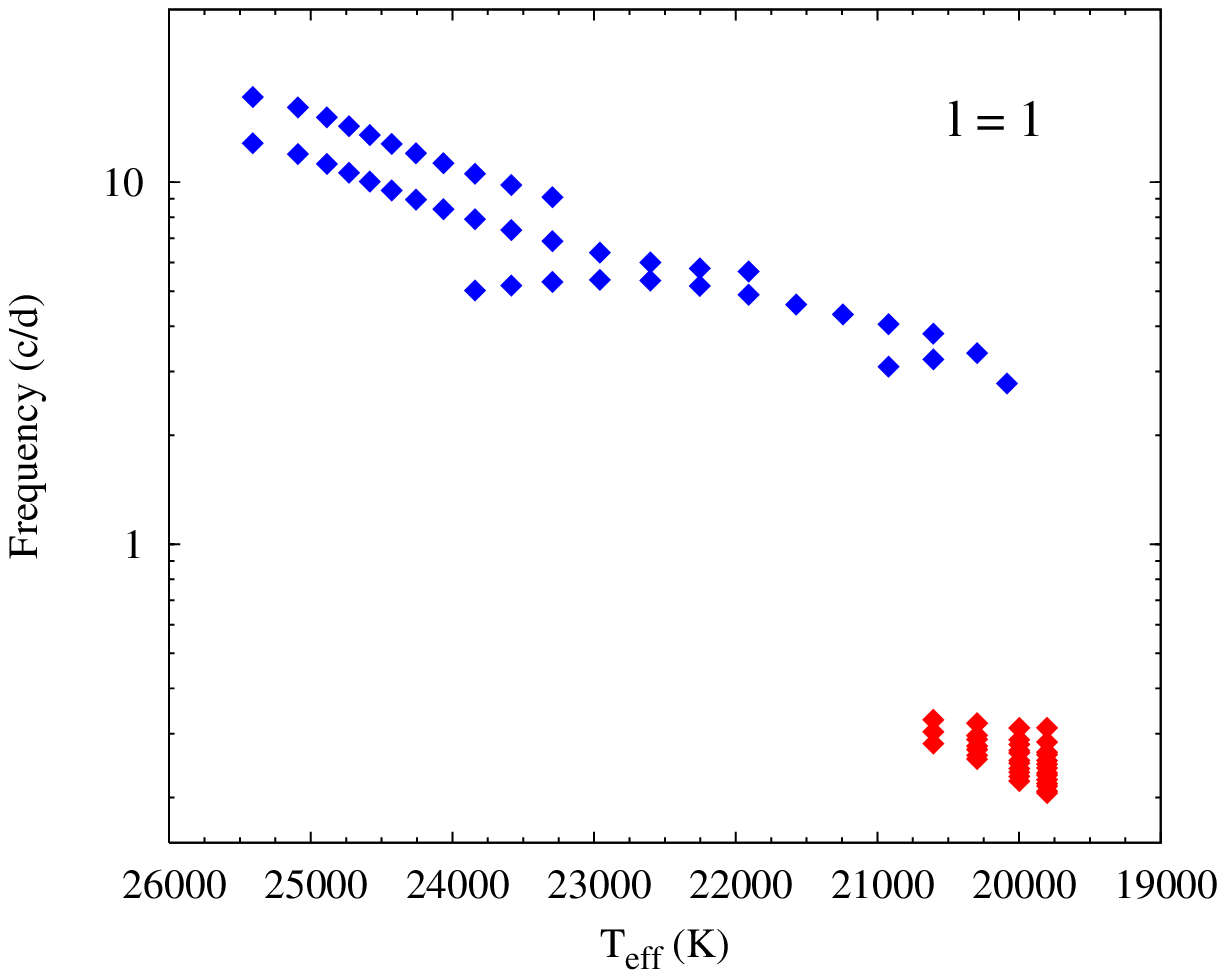}
\end{minipage} 
\begin{minipage}{37pc}
\caption{\label{evoll1l2}
Frequencies of unstable $\ell=2$ (left) and $\ell=1$ (right) modes of
$\beta$~Cep-type (blue) and SPB-type (red) as a function of $T_{\rm eff}$, along the evolution
sequence of 10~M$_\odot$ from ZAMS to TAMS with OP opacities.}
\end{minipage} 
\end{figure}

Recent observations show the existence of hybrid B stars with at the same time oscillations
of $\beta$~Cep type ($\nu\approx 4-8\;{\rm d}^{-1}$) and high order g-modes of SPB type
($\nu\approx 0.2-1.2\;{\rm d}^{-1}$). The clearest cases are 19~Mon \cite{balona}, 
$\nu$~Eri  \cite{handler04} and 12~Lac \cite{handler06}. Another very promising star is
HD~180642 (a primary target of COROT). Several other cases exist such as $\gamma$~Peg \cite{chapellier},
but they must be considered with caution because of binarity
or because of the low S/N and the risks of aliases, asking for more 
precise observations.  

From a theoretical point of view, \cite{pam}, \cite{miglio} and \cite{migliocoast} showed that with OP opacities
(\cite{seaton96}, \cite{seaton05}), 
the instability strips of $\beta$~Cep and SPBs intersect, so that theoretical models with 
both unstable p-modes and high order g-modes exist. 
An illustration is given in Fig.~\ref{evoll1l2} for 10~M$_\odot$ models.

\begin{figure}[h]
\begin{minipage}{19pc}
\includegraphics[width=19pc]{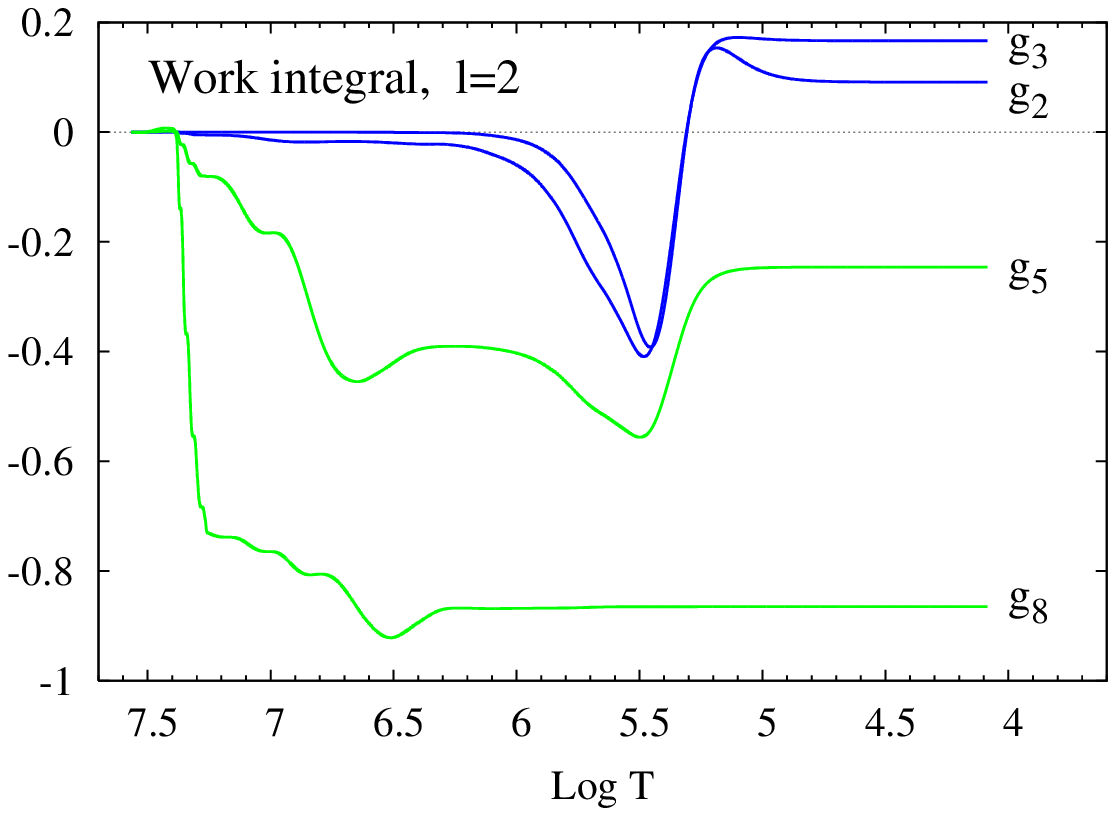}

\vspace{-0.3cm}
\includegraphics[width=19pc]{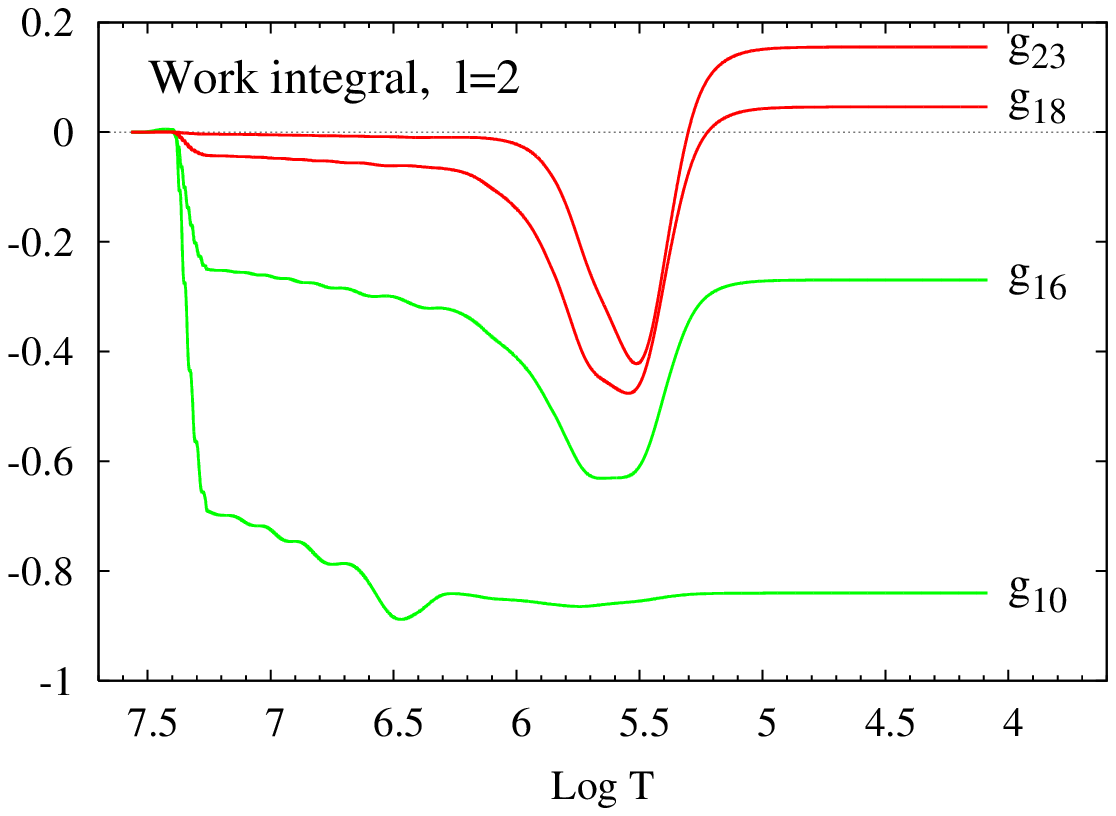}
\end{minipage}\hspace{0pc}
\begin{minipage}{19pc}
\includegraphics[width=19pc]{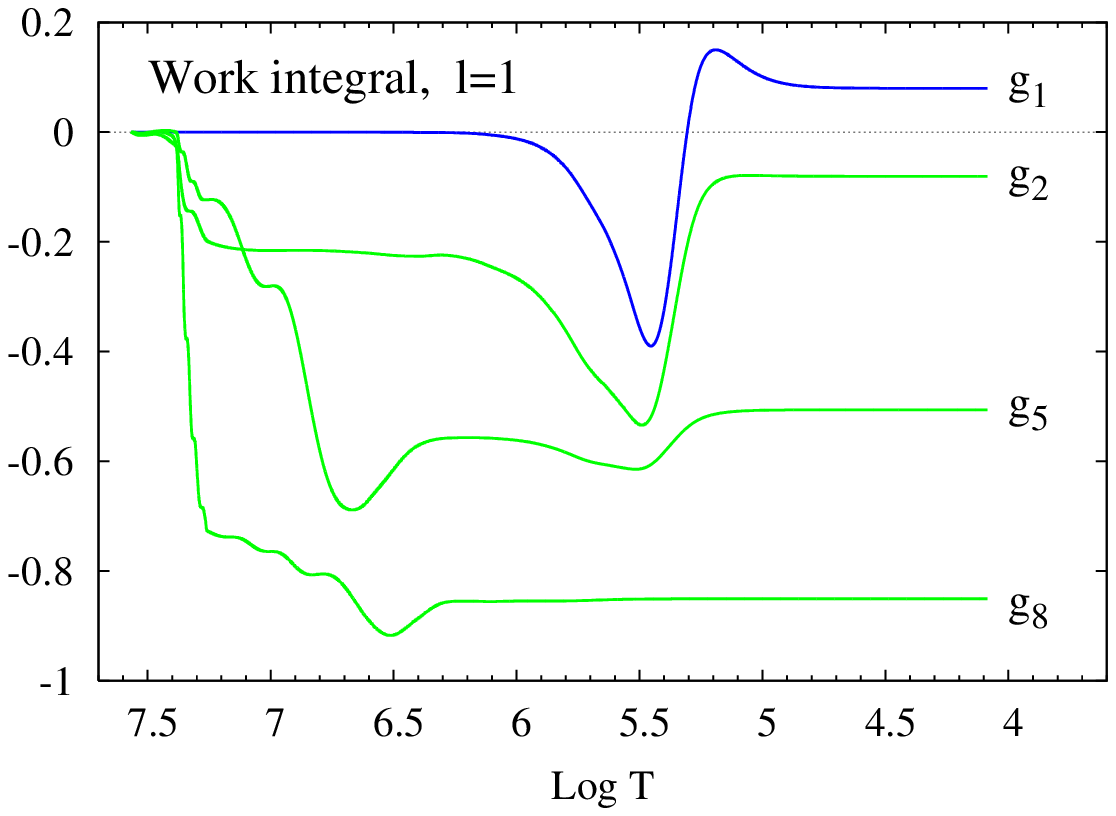}

\vspace{-0.3cm}
\includegraphics[width=19pc]{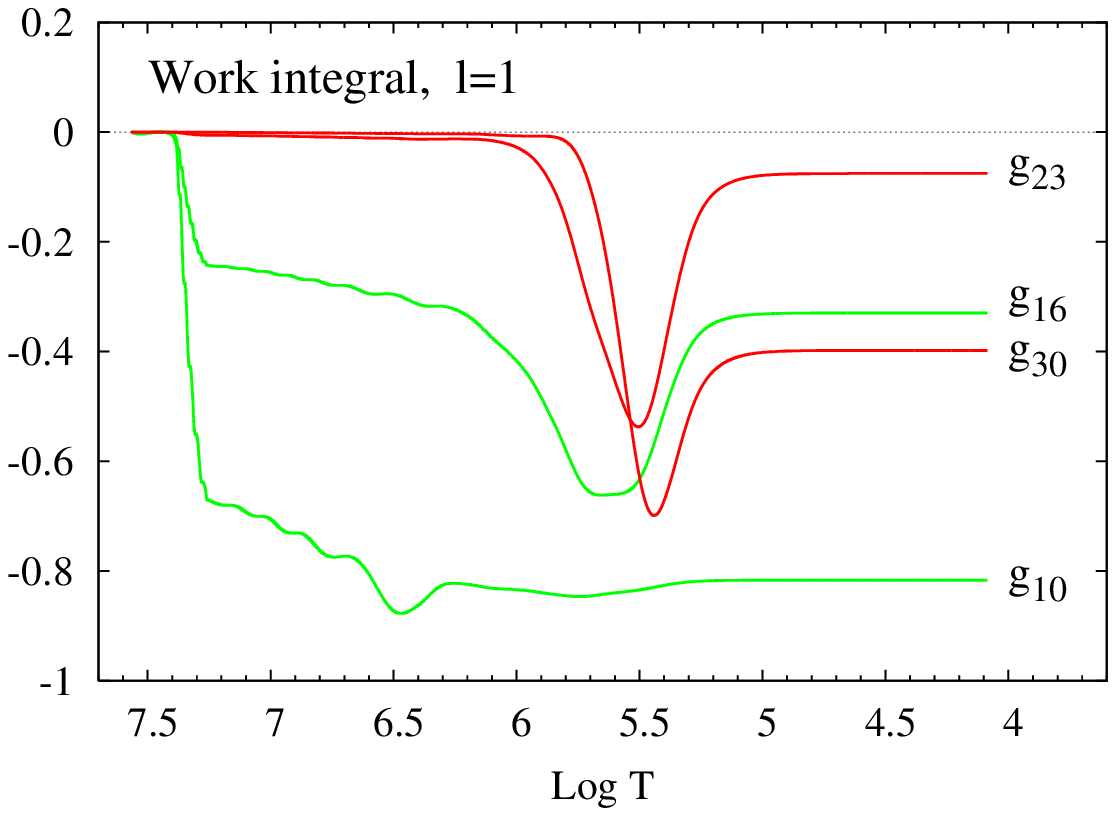}
\end{minipage}
\begin{minipage}{37pc}
\caption{\label{work}Work integrals $\int_0^m {\rm d}W/\int_0^M |{\rm d} W|$ for different $\ell=2$ modes (left) and $\ell=1$ modes (right) of a 10~M$_\odot$ model with $T_{\rm eff}=21245$~K,
$\log(L/L_\odot)=4.06$ and OP opacities.}
\end{minipage}
\end{figure}

For the structure models of this study, we used the stellar evolution code CLES \cite{scuflaire}, 
with by default the following physical prescriptions: the new OP opacities \cite{seaton05},
the OPAL2001 equation of state  \cite{rogerset}, the mixture of elements by \cite{asplund}, 
with the enhancement of Ne proposed by \cite{cunha}.

\section{Driving and damping regions}

To identify the driving and damping regions, we give in Fig.~\ref{work} the work integrals for different modes of a well chosen 10~M$_\odot$ model. Left panels correspond to $\ell=2$ modes
and right panels to $\ell=1$ modes.
Regions where $W$ increases (resp. decreases) outwards have a driving (resp. damping) effect. We analize first the $\ell=2$ modes (left). Unstable modes of $\beta$~Cep type (low radial order mixed modes) are given in blue in the top left panel and unstable modes of SPB type (high radial order g-modes) are given in red in the bottom left panel. For these two types of unstable modes, we see clearly the significant driving around the IOB at $\log T\approx 5.3$. But g-modes of intermediate radial order (green curves) are stable because of a significant damping in the very deep layers just above the convective core. This damping is a classical radiative damping mechanism. The short-wavelength oscillations in the g-mode cavity due to the large values of the Brunt-V\"ais\"al\"a frequency lead to very large values of the derivatives
of the eigenfunctions (${\rm d}/{\rm d} r (f(r)\sin(\int k {\rm d}r))\simeq k\:f(r)\cos(\int k {\rm d}r)$ with $k\simeq\sqrt{\ell(\ell+1)}\sigma^{-1}\int N/r \:{\rm d}r>>0$). In particular, the temperature gradient variations due to the oscillations are very large in this region; it is easily seen that they lead to significant loss of heat during the hot phase (see Eq.~7 of \cite{dupret05}), which always damps the modes. But why this significant damping occurs for some g-modes and not for others~?

\begin{figure}[h]
\begin{minipage}{19pc}
\includegraphics[width=19pc]{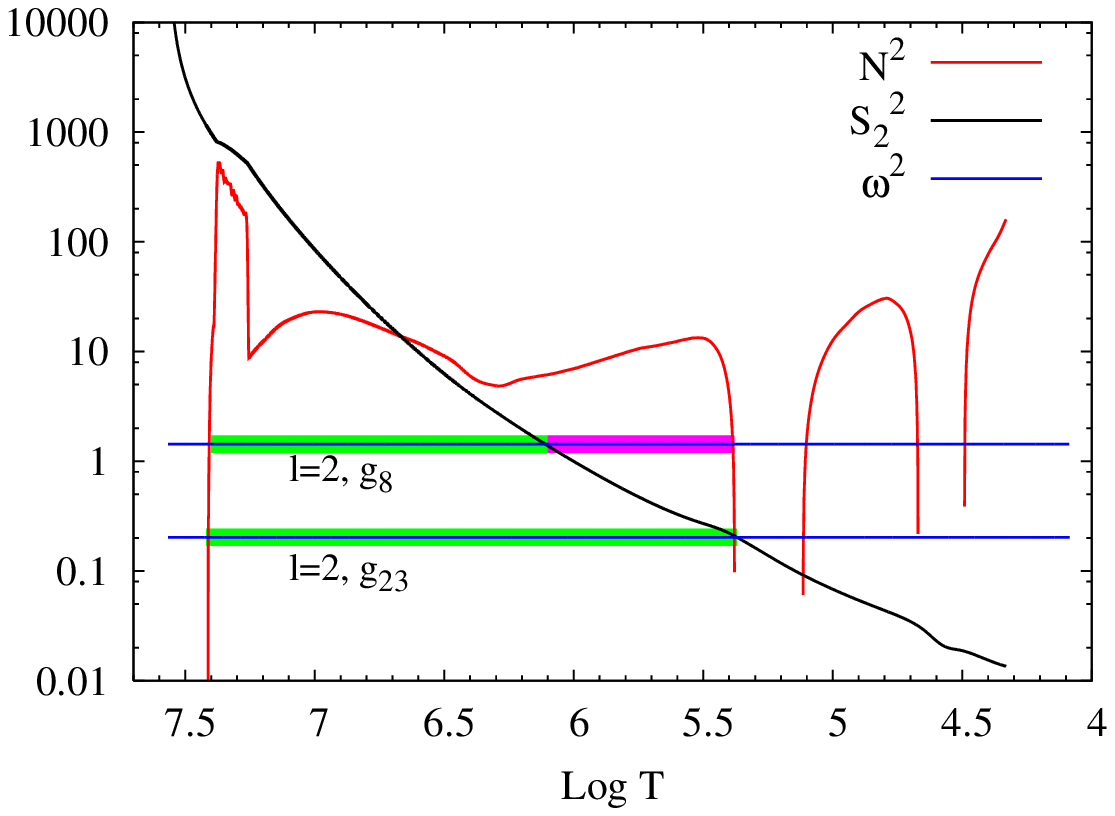}
\end{minipage}\hspace{0pc}%
\begin{minipage}{19pc}
\includegraphics[width=19pc]{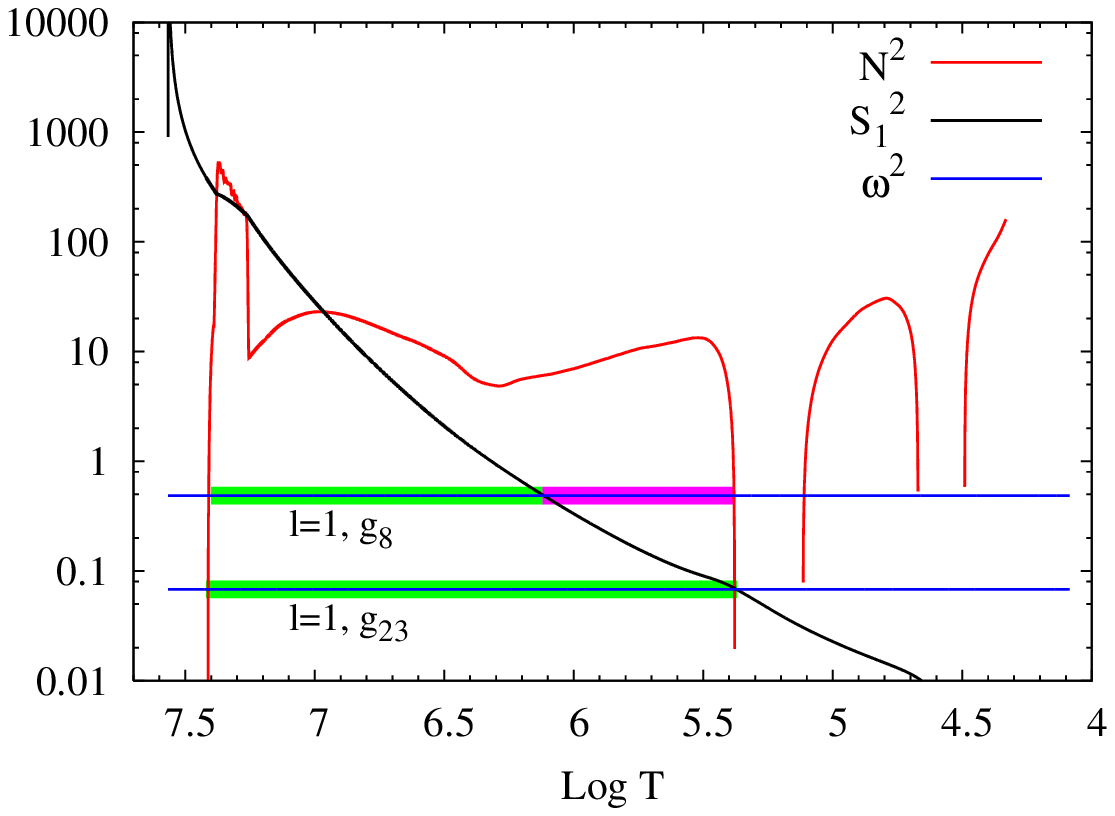}
\end{minipage} 
\begin{minipage}{37pc}
\caption{\label{vaisB}
Square of the dimensionless Brunt-V\"ais\"al\"a frequency $N^2\:R^3/(GM)$ (red),
Lamb frequency $S_\ell^2\:R^3/(GM)$ (black) and angular pulsation frequencies 
$\omega^2=\sigma^2\:R^3/(GM)$ (horizontal blue lines), for $\ell=2$ modes (left) and $\ell=1$ modes
(right).}
\end{minipage} 
\end{figure}

\begin{figure}[h]
\begin{minipage}{19pc}
\includegraphics[width=19pc]{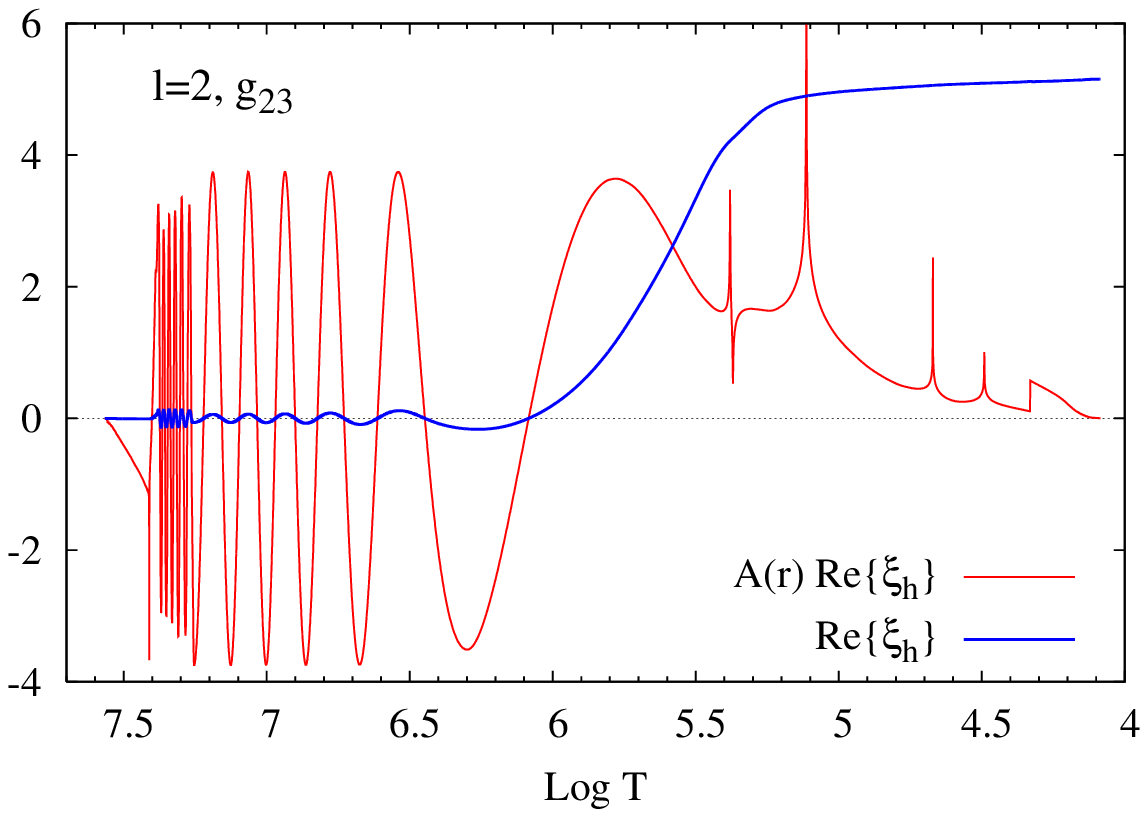}
\end{minipage}\hspace{0pc}%
\begin{minipage}{19pc}
\includegraphics[width=19pc]{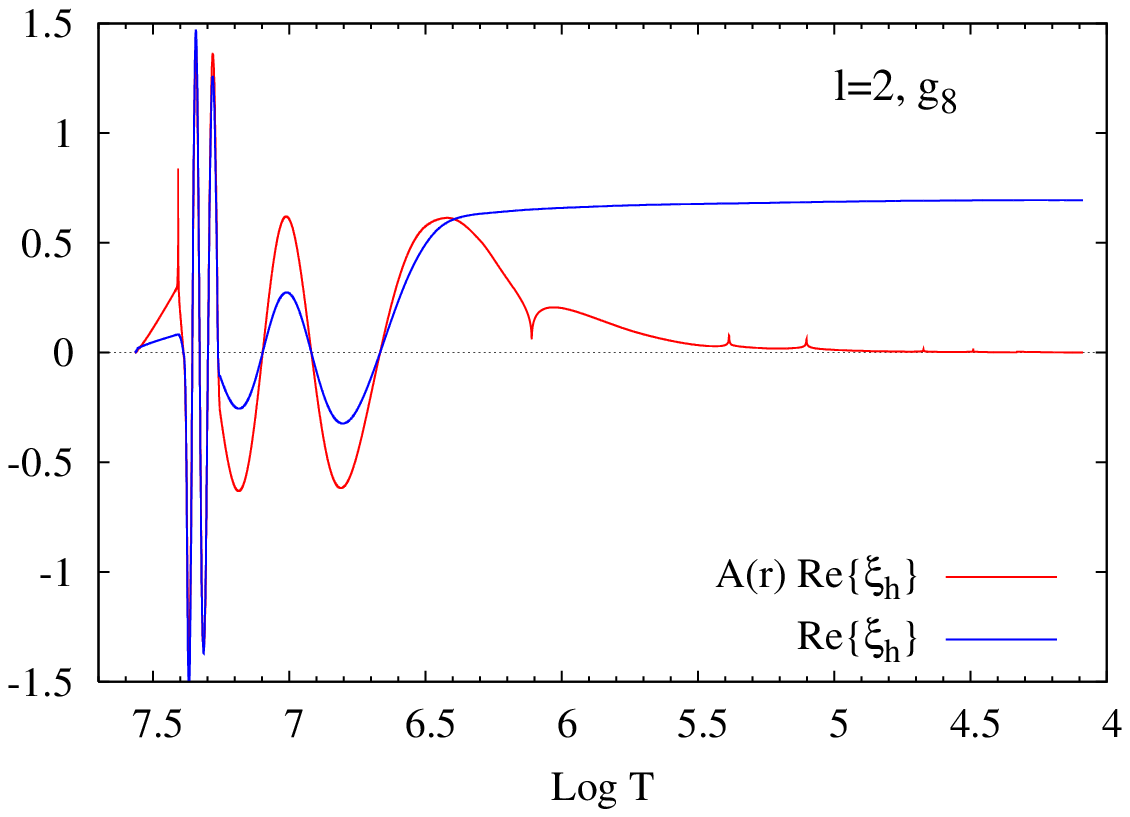}
\end{minipage} 
\begin{minipage}{37pc}
\caption{\label{xih} Real parts of the transversal displacement $\xi_h$ (blue) and \\$A(r)\xi_h\:=\:\rho^{1/2} r^{2} c^{-1/2} |S_\ell^2/\sigma^2-1|^{1/4}\:|N^2/\sigma^2-1|^{-1/4}\;\xi_h$ (red), for the modes $\ell=2$, g$_{23}$ (left) and $\ell=2$, g$_{8}$ (right).}
\end{minipage} 
\end{figure}

\section{Propagation and evanescent regions}

Changes in the respective weights of the driving and damping regions originate in the shape of the eigenfunctions. To understand this, classical propagation diagrams are very useful, as shown in Fig.~\ref{vaisB} for the 10 M$_\odot$ model considered previously. According to the asymptotic theory, the transversal component of the displacement (dominating for g-modes) behaves as:
\begin{equation}
\xi_h\;\propto\;\frac{c^{1/2}}{\rho^{1/2}\: r^{2}}\,\left|\frac{N^2/\sigma^2-1}{S_\ell^2/\sigma^2-1}\right|^{1/4}F\left(\int K(r)\:{\rm d}r\right)
\:\approx \:\rho^{-1/2} r^{-3/2}\:F\left(\frac{\sqrt{\ell(\ell+1)}}{\sigma}
\int N/r {\rm d}r\right)\;,
\end{equation}
where the function $F$ is a cosine of constant amplitude in the propagation regions (green in Fig.~\ref{vaisB}), and an exponential decreasing outwards in the evanescent regions (magenta in Fig.~\ref{vaisB}). Fig.~\ref{xih} (left panel)
shows that this formula applies well for the mode $\ell=2$, g$_{23}$. The propagation region is very large for this mode going up to the IOB,  the change of amplitude of $\xi_h$ is dominated there by the factor $\rho^{-1/2}$ which leads to values much smaller near the center compared to the surface. Hence, the deep radiative damping is small compared to the $\kappa$-driving in the IOB and this mode is unstable. The right panel of Fig.~\ref{xih} shows the same eigenfunctions for the mode $\ell=2$, g$_8$. Now the propagation region is smaller with a larger evanescent region above it (see Fig.~\ref{vaisB}). The corresponding exponential decrease leads to amplitudes much smaller around the IOB compared to the deep layers. Hence the deep radiative damping overwhelms the $\kappa$-driving and this mode is stable. 

\section{Dependence with the spherical degree $\ell$}

In the right panels of Fig.~\ref{work} the work integrals for the same model as before are given, but now for $\ell=1$ modes. We see in this case that the $\beta$~Cep-type mixed mode $\ell=1$, g$_1$ is predicted to be unstable (blue, top right panel), but all $\ell=1$ high radial order g-modes of SPB type are predicted to be stable, contrary to the $\ell=2$ case. 
This can be understood as follows.
At high frequencies, the eigenfunctions do not depend significantly on $\ell$. Hence the range of excited modes of $\beta$~Cep type (as a function of the frequency) remains essentially the same for different $\ell$ (blue diamonds in Fig.~\ref{evoll1l2} and blue curves in the top panels of Fig.~\ref{work}). At lower frequencies, we have seen that the respective weight of the damping and driving regions is determined by the size of the propagation and evanescent regions. As $S_\ell^2 \propto \ell (\ell+1)$, g-modes having the same values of $\ell (\ell+1)/\sigma^2$ have the same propagation regions. This corresponds to a lower frequency for $\ell=1$ compared to $\ell=2$ modes (compare the left and right panels of Fig.~\ref{vaisB}). Hence the critical frequency at which the upper boundary of the propagation region reachs the IOB is smaller for $\ell=1$ modes. At this smaller frequency, the transition region where the pulsation period is of the same order as the thermal relaxation time is deeper in the star, around the hot side of the IOB; so the damping effect of this region is larger than the driving of the more superficial layers (red curves in the right bottom panel of Fig.~\ref{work}) and all high order g-modes are stable. 

\section{OP versus OPAL}

\begin{figure}[h]
\begin{minipage}{18.5pc}
\includegraphics[width=18.5pc]{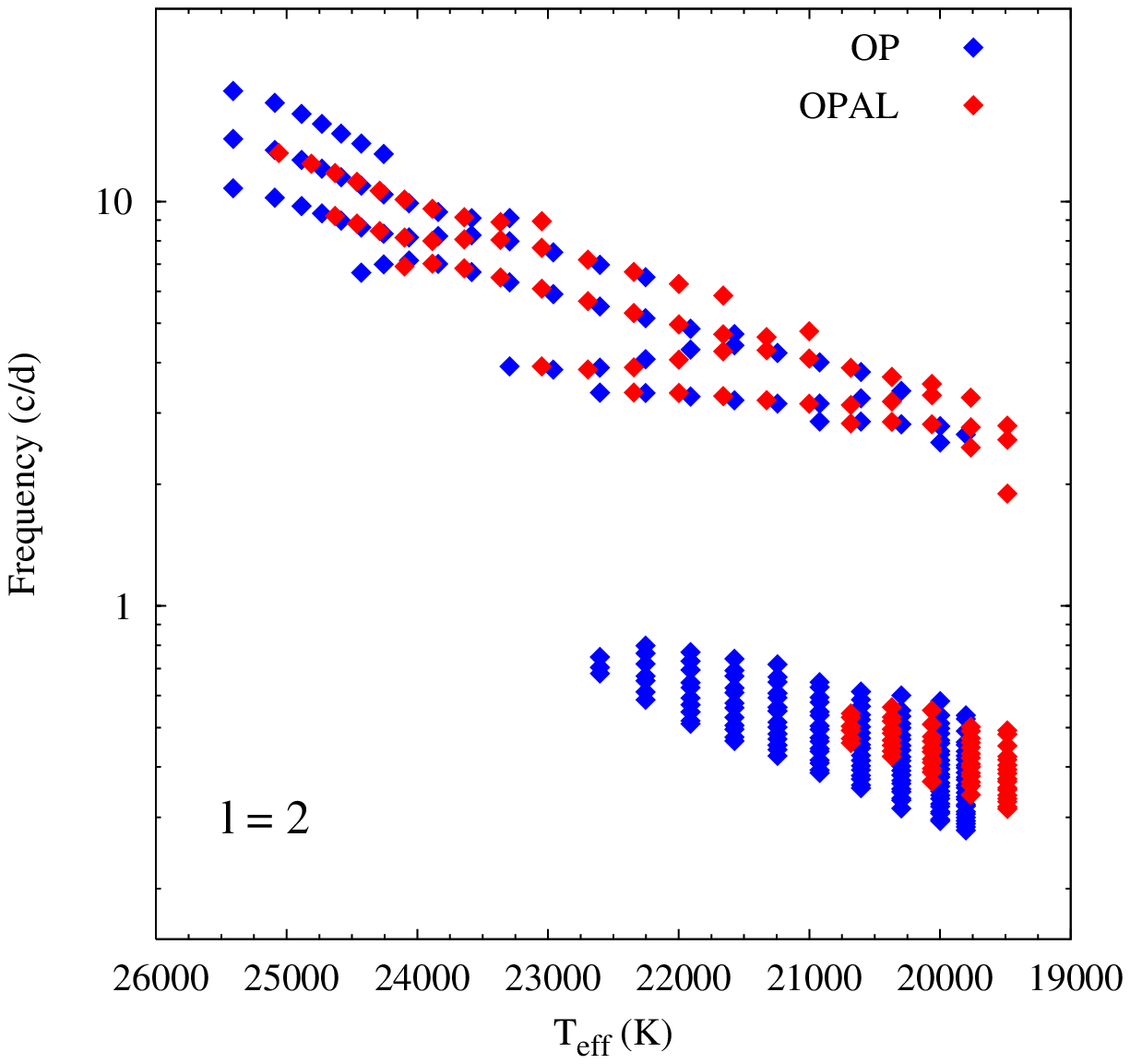}
\caption{\label{evolopopal}
Frequencies of unstable $\ell=2$ modes as a function of $T_{\rm eff}$, along the evolution
sequence of 10~M$_\odot$ from ZAMS to TAMS, using OP (blue) and OPAL (red) opacities.}
\end{minipage}\hspace{1pc}
 \begin{minipage}{18.3pc}
\includegraphics[width=18.3pc]{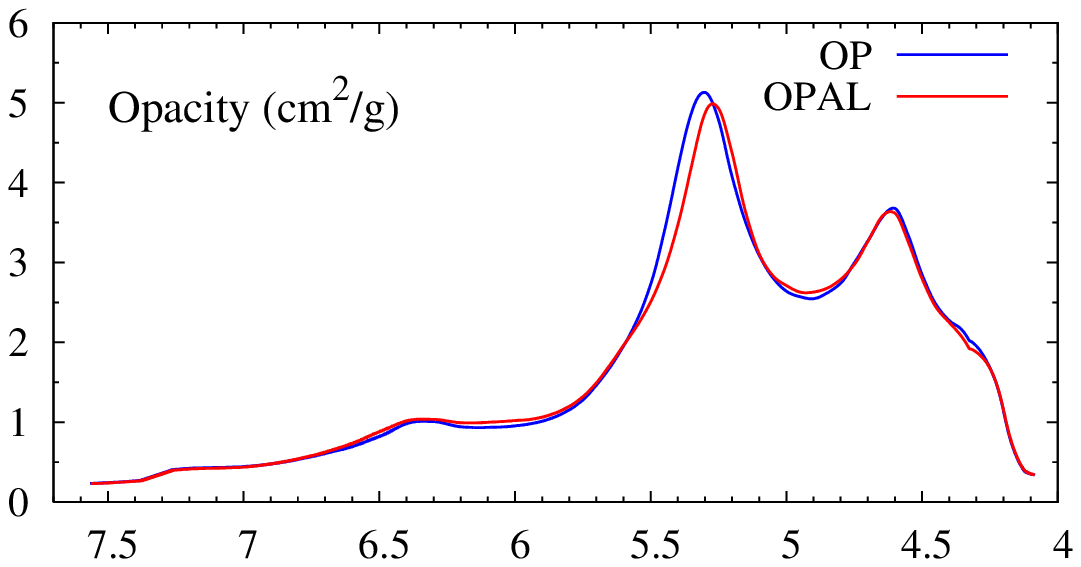}

\vspace{-0.3cm}
\includegraphics[width=18.3pc]{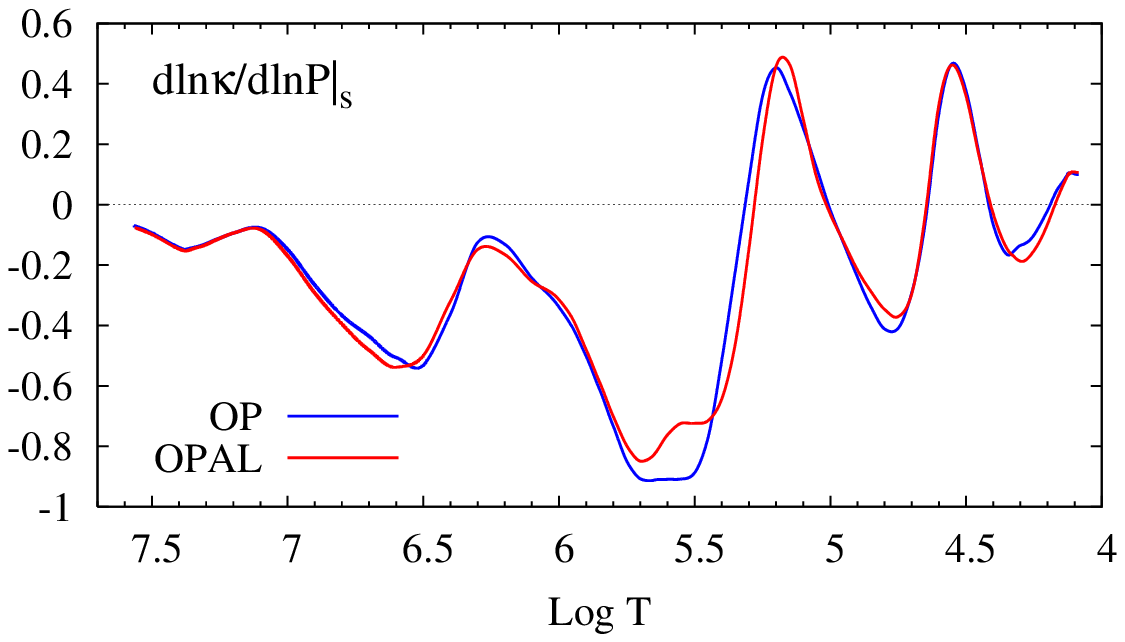}
\caption{\label{kappa}
Opacities (top) and $\partial\ln\kappa/\partial\ln P|_s$ (bottom) obtained
with OP (blue) and OPAL (red) opacity tables, 10~M$_\odot$ models with $T_{\rm eff}=21245$~K and
$\log(L/L_\odot)=4.06$.}
\end{minipage}
\end{figure}

\begin{figure}[h]
\begin{minipage}{19pc}
\includegraphics[width=19pc]{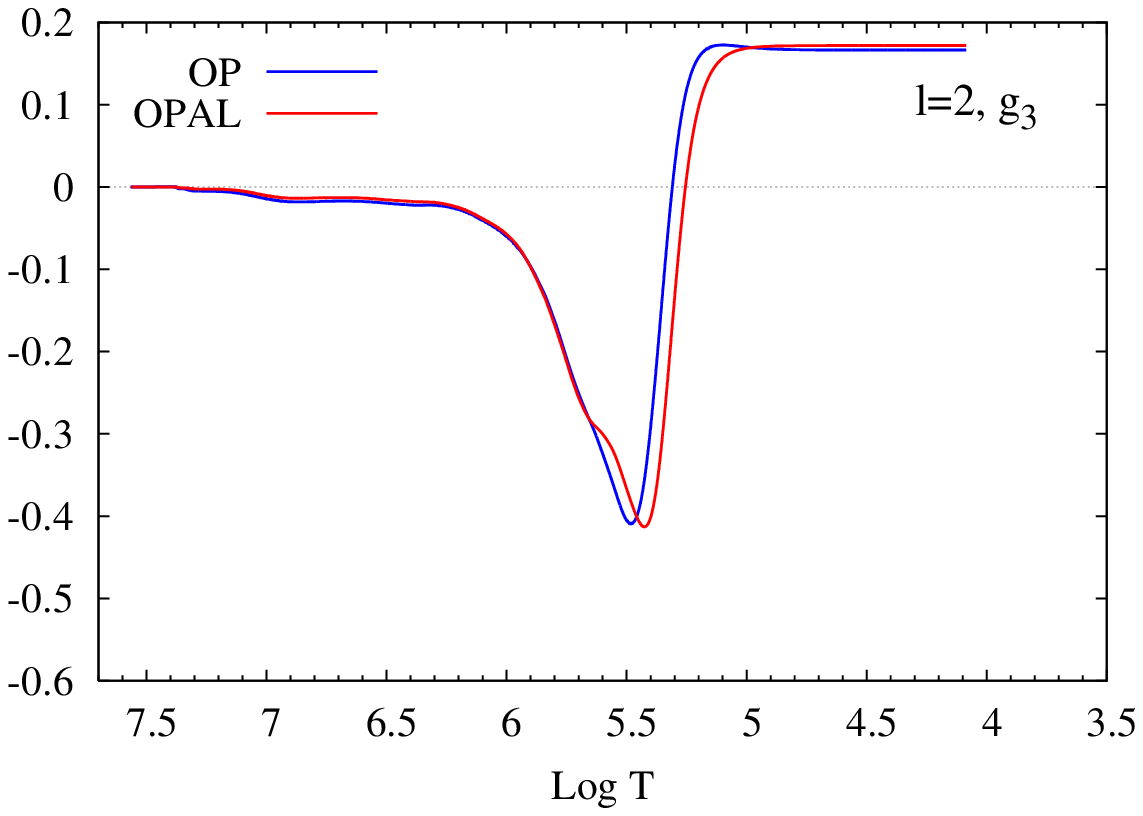}
\end{minipage}\hspace{0pc}%
\begin{minipage}{19pc}
\includegraphics[width=19pc]{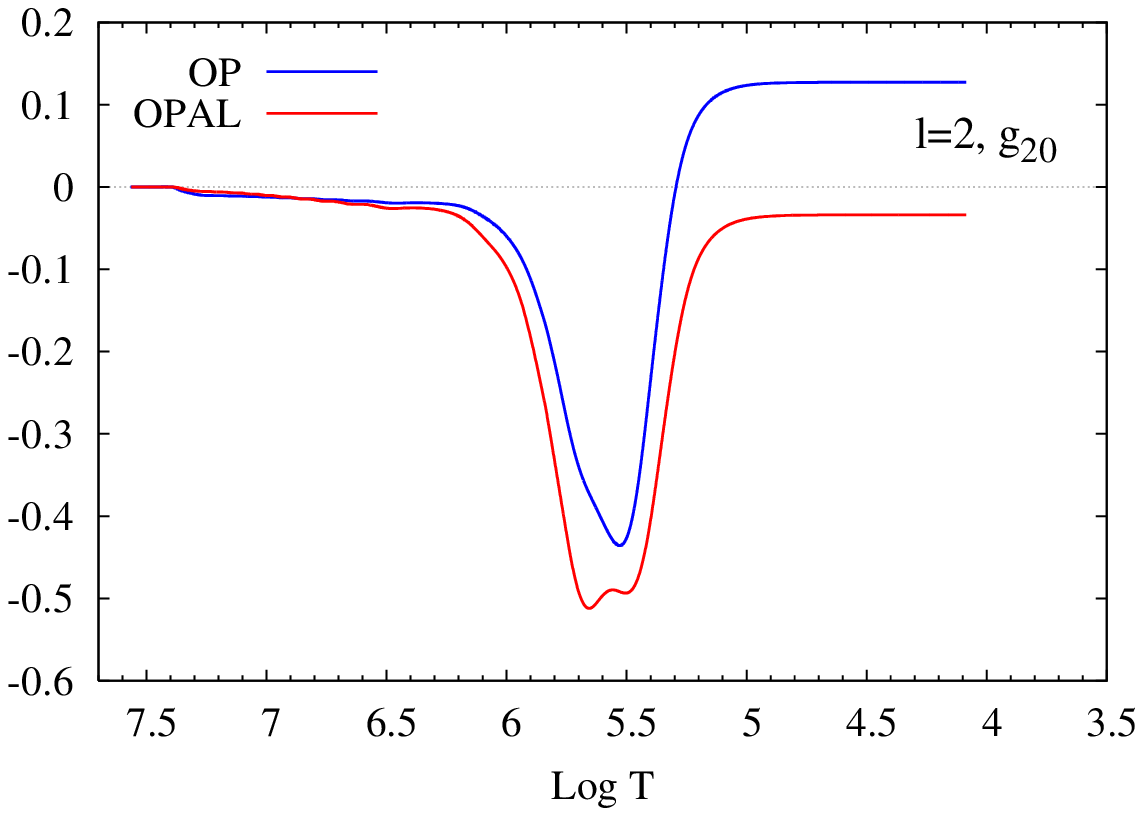}
\end{minipage} 
\begin{minipage}{37pc}
\caption{\label{workopopal}Work integrals for the mode $\ell=2$, g$_3$ (left) and $\ell=2$, 
g$_{20}$ (right) of the same models as Fig.~\ref{kappa}, using OP (blue) and OPAL (red) opacities.}
\end{minipage} 
\end{figure}

\cite{miglio} showed that many more hybrid $\beta$~Cep-SPB are predicted when OP opacity tables
(\cite{seaton05}) are used compared with OPAL (\cite{iglesias}). This can be seen in 
Fig.~\ref{evolopopal} where we compare the range of excited modes obtained in the two cases for
 10~M$_\odot$ models.
The comparison of the OP and OPAL opacities and their derivative 
$\partial\ln\kappa/\partial\ln P|_s$ presented in Fig.~\ref{kappa} 
shows significant differences around the hot wing of the IOB.
Fig.~\ref{workopopal} compares the work integrals obtained with OP and OPAL for a typical 
$\beta$~Cep-type mode (left) and SPB-type mode (right). The results are very close
for $\beta$~Cep-type modes because the opacities are essentially the same with OP and OPAL 
in the superficial layers ($\log T< 5.2$) where p-modes have significant amplitudes.
But the results are different for the high order g-modes: with OP they are excited while
most are stable with OPAL. To understand this, I recall that two things are required for an
efficient driving of the g-modes. First, the transition region where the pulsation period 
is of the same order as the thermal relaxation time must coincides with the IOB. And 
second, this period must be large enough so that the corresponding evanescent region is 
negligible. These two requirements are fulfilled with OP; but with OPAL the period at 
which the transition region and the IOB coincide is too small, leading to a too large 
evanescent region. 
This explains why
the driving of high order g-modes (probing these deep layers) is more efficient
with OP than with OPAL, and more hybrid models are predicted with OP.


\newpage

\section{Hybrid $\delta$~Sct-$\gamma$ Dor stars}

\begin{figure}[h]
\begin{minipage}{19pc}
\includegraphics[width=19pc]{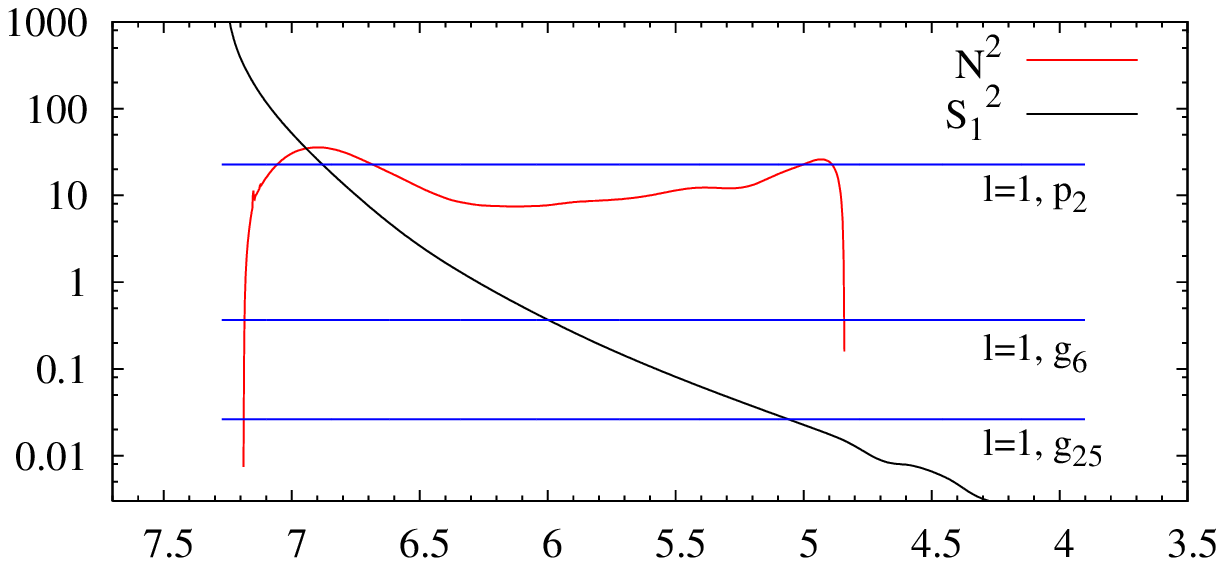}

\vspace{-0.3cm}
\includegraphics[width=19pc]{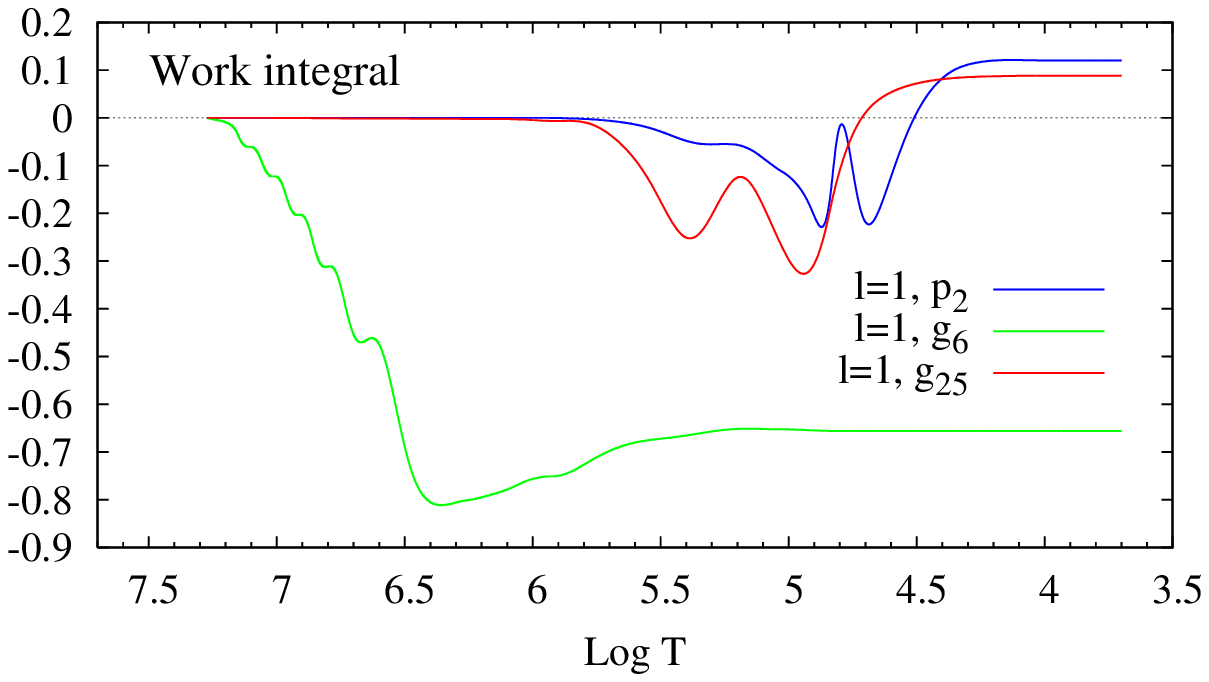}
\caption{\label{workdsgd}
Propagation diagram (top) and 
work integrals (bottom) for the modes $\ell=1$, p$_2$, g$_6$ and g$_{25}$ of a 
young 1.55~M$_\odot$ $\delta$~Sct-$\gamma$~Dor model.}
\end{minipage}\hspace{1pc}
\begin{minipage}{18pc}
\includegraphics[width=18pc]{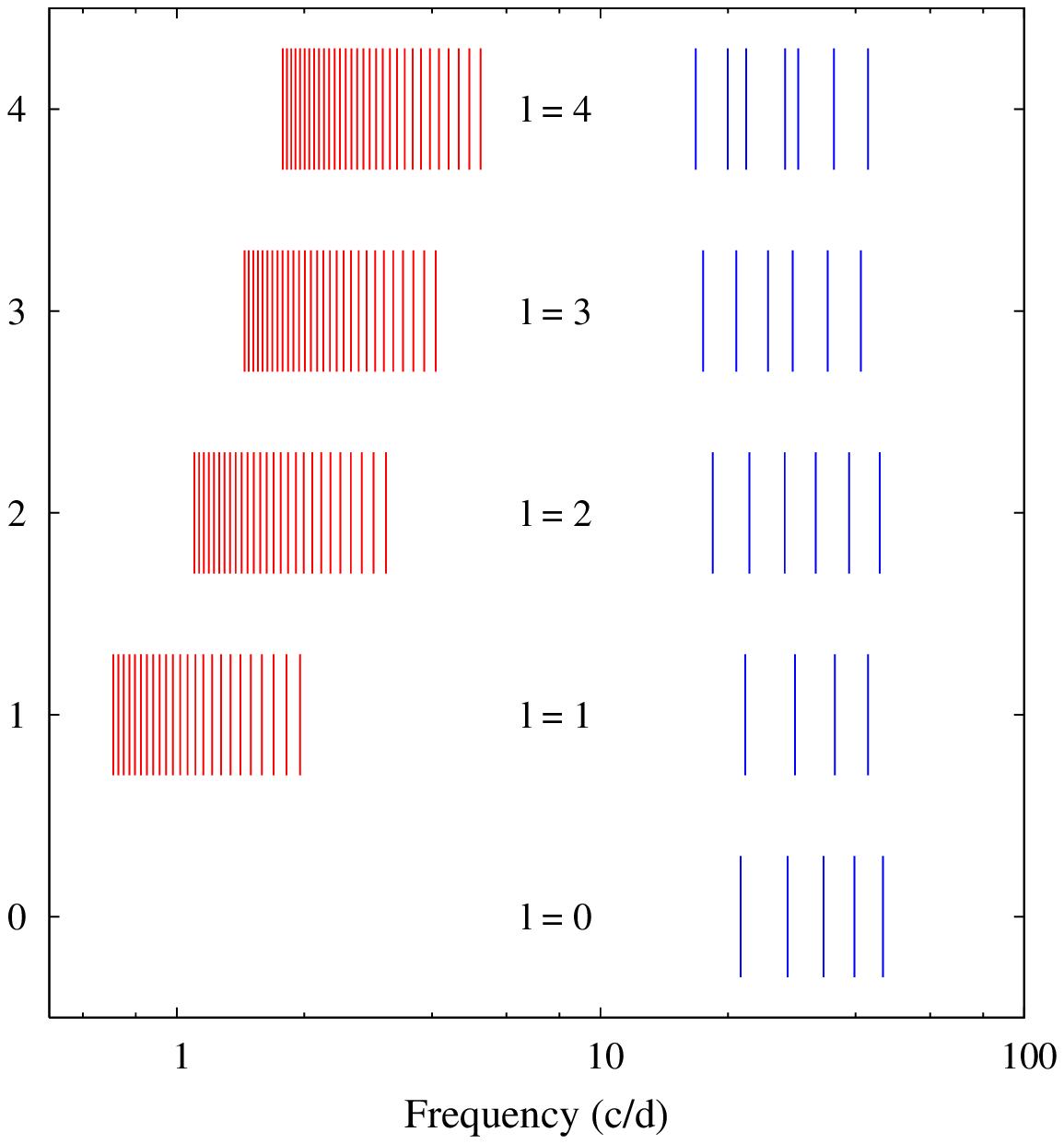}
\caption{\label{dsgdis}Frequencies of unstable $\ell=1-4$ modes of $\gamma$~Dor type (red) and $\delta$~Sct type (blue), for the same model as Fig.~\ref{workdsgd}. 
}
\end{minipage} 
\end{figure}


Lower along the main sequence, hybrid models with both unstable $\delta$~Sct-type mixed modes and 
$\gamma$~Dor-type g-modes are also predicted (\cite{dupret05}, \cite{grigahcene06}) 
The observational search of such hybrid models was less conclusive than in B 
stars but some cases were found: the Am star HD~8801 \cite{henry} and the binary HD~209295
\cite{handler02}. The driving of these stars is very different from the one of B stars, being complicated by the 
time-dependent coherent interaction with convection for cool models \cite{dupret05}. However, we point out here
that some aspects of the damping mechanisms remain very similar.

In Fig.~\ref{workdsgd}, we give a propagation diagram (top, dimensionless frequencies) and the work integrals (bottom)
for the modes $\ell=1$, p$_2$, g$_6$ and g$_{25}$ of a young  1.55~M$_\odot$ $\delta$~Sct-$\gamma$~Dor hybrid model 
with  $T_{\rm eff}=7330$~K, $\log(L/L_\odot)=0.746$ and $\log g=4.295$. This model is computed with OPAL opacities and $Z=0.02$.
We used for the non-adiabatic computation the Time-Dependent Convection (TDC) treatment of \cite{grigahcene}.
The mode $\ell=1$, p$_2$ (blue, $\nu=28.8$ c/d) is a typical unstable mode of $\delta$~Sct type; TDC plays a central role
in its driving and damping. The mode
$\ell=1$, g$_{25}$ (red, $\nu=0.98$ c/d) is a typical unstable mode of $\gamma$~Dor type; the small driving
at $\log T\simeq 5.3$ comes from a $\kappa$-mechanism in the IOB and the significant driving at 
$\log T\simeq 4.8$ is a flux blocking mechanism operating a the base of the convective envelope.
The mode $\ell=1$, g$_6$ (green, $\nu=3.66$ c/d) is a stable mode which is significantly damped in the deep radiative 
g-mode cavity.
We note in the top panel of Fig.~\ref{workdsgd} that the sizes of the propagation and evanescent regions for these modes
are similar to the case of B stars considered above. 
Hence, the explanation of the stable gap between the $\delta$~Sct and $\gamma$~Dor unstable modes is the same as 
before: these g-modes of intermediate radial order have a large evanescent region; hence the amplitudes are large
in the deep g-mode cavity, and the radiative damping occurring there is much larger than the driving of the 
superficial layers. We note that the model considered here is very young, without region of variable molecular weight above the convective core. Hence, no bump of the  Brunt-V\"ais\"al\"a
frequency is present there.

In Fig.~\ref{dsgdis}, all the frequencies of unstable $\ell=1-4$ modes are given for the same model as before.
We note that the stable gap between the $\delta$~Sct and $\gamma$~Dor unstable modes decreases as $\ell$ increases.
This is simply due to the fact that $S_\ell^2 \propto \ell(\ell+1)$, so that the size of the propagation cavity of 
g-modes increases with $\ell$ for a given frequency.





\section{Conclusions}

Theoretical models and observations indicate the existence of hybrid models having at the same time 
unstable pressure and gravity modes.
Concerning first the driving mechanism of hybrid $\beta$~Cep-SPB models we have shown, in addition to its 
already known sensitivity to the iron abundance, that the characteristics of the propagation and evanescent regions 
also play a major role in this context, determining the extension of the stable gap in the frequency domain 
between the unstable low order pressure and high order gravity modes. The same phenomenon occurs for the  
 hybrid $\delta$~Sct-$\gamma$~Dor pulsators. Hence, the comparison between the observed and theoretical ranges
of unstable modes allows us to constrain the Brunt-V\"ais\"al\"a and Lamb frequencies in the very deep interior
of these stars.

\smallskip





\end{document}